ARTICLE OPEN

# Probing the local nature of excitons and plasmons in few-layer MoS$_2$

Hannah Catherine Nerl[1], Kirsten Trøstrup Winther[2], Fredrik S. Hage[3], Kristian Sommer Thygesen[2], Lothar Houben[4], Claudia Backes[1], Jonathan N. Coleman[1], Quentin M. Ramasse[3] and Valeria Nicolosi[1,5]

Excitons and plasmons are the two most fundamental types of collective electronic excitations occurring in solids. Traditionally, they have been studied separately using bulk techniques that probe their average energetic structure over large spatial regions. However, as the dimensions of materials and devices continue to shrink, it becomes crucial to understand how these excitations depend on local variations in the crystal- and chemical structure on the atomic scale. Here, we use monochromated low-loss scanning-transmission-electron-microscopy electron-energy-loss spectroscopy, providing the best simultaneous energy and spatial resolution achieved to-date to unravel the full set of electronic excitations in few-layer MoS$_2$ nanosheets over a wide energy range. Using first-principles, many-body calculations we confirm the excitonic nature of the peaks at ~ 2 and ~ 3 eV in the experimental electron-energy-loss spectrum and the plasmonic nature of higher energy-loss peaks. We also rationalise the non-trivial dependence of the electron-energy-loss spectrum on beam and sample geometry such as the number of atomic layers and distance to steps and edges. Moreover, we show that the excitonic features are dominated by the long wavelength ($q = 0$) components of the probing field, while the plasmonic features are sensitive to a much broader range of q-vectors, indicating a qualitative difference in the spatial character of the two types of collective excitations. Our work provides a template protocol for mapping the local nature of electronic excitations that open new possibilities for studying photo-absorption and energy transfer processes on a nanometer scale.

## INTRODUCTION

Transition-metal dichalcogenides (TMDs) form a large and versatile class of two-dimensional (2D) layered materials.[1–6] In particular, the semi-conductor MoS$_2$ has shown great promise for diverse applications including catalysis,[7] battery electrodes,[8] field effect transistors[9] and solar cells.[10, 11] One highly attracting feature of the TMDs is that their electronic properties can be tuned by varying the number of layers in the film. For example, the energy of the lowest optical transition in MoS$_2$ increases from 1.2 eV (ref. [12]) in bulk to 1.9 eV (ref. [13]) for a monolayer with an associated strong increase in photoluminescence[14] that has been attributed to the transition from indirect to direct band gap.[15–18] In addition, MoS$_2$ exhibits strong spin–orbit interaction that leads to a coupling of the spin and valley degrees of freedom in monolayer samples with broken inversion symmetry.

Understanding how the local structure and chemistry of the 2D materials influence their electronic properties is crucial in order to harness the full potential of these novel materials. For most applications in photonics and electronics, a smooth and uniform surface is essential. However, in practice, local variations in crystal structure, film thickness and chemical composition are prevalent. These variations are bound to influence the electronic properties of the materials such as the (local) band gap, the energy and lifetimes of various quasiparticles and collective excitations. Hence, there is a need to fully characterise the physical and chemical properties of nanostructured materials from the bottom up i.e., at the level of individual atoms, and to map out the opto-electronic properties where they happen. Previous studies have relied on bulk techniques to study these properties of 2D materials. However, the effect of local variations in structure such as flake edges or number of layers has been difficult to analyse from such measurements due to lack of spatial resolution.

In this work, we use low-loss scanning-transmission-electron-microscopy electron-energy-loss (LL-STEM-EEL) spectroscopy to study the electronic excitation spectrum of few-layer MoS$_2$ in the near-infrared/visible/ultraviolet spectral ranges. Two main types of energy-loss features are expected in this energy range, and in our experimental conditions: the most prominent losses are due to plasmonic excitations (or plasmons), arising from the collective excitation of the material's valence electrons (for more information, see for instance a recent review, and references therein)[19]; additionally, the electronic structure of MoS$_2$ is known to support excitonic transitions (or excitons), arising from the interaction between an electron excited by the beam to an empty state in the material's electronic structure and the 'hole' left behind in the valence band. Compared to conventional techniques for measuring optical properties, LL-STEM-EEL spectroscopy offers the unique combination of spatial and energy resolution, which opens up new possibilities for studying the local optical properties with an unprecedented energy resolution over a wide spectral range.

[1]CRANN & AMBER and School of Physics, Trinity College Dublin, Dublin 2, Ireland; [2]Center for Nanostructured Graphene (CNG), Department of Physics, Technical University of Denmark, Fysikvej 2800 Kongens Lyngby, Denmark; [3]SuperSTEM Laboratory, STFC Daresbury Campus, Keckwick Lane, Daresbury WA4 4AD, UK; [4]Ernst Ruska Center for Microscopy and Spectroscopy with Electrons, Research Center, Juelich, Germany & Weizmann Institute of Science, Rehovot, Israel and [5]School of Chemistry, Trinity College Dublin, Dublin 2, Ireland
Correspondence: Hannah Catherine Nerl (nerlh@tcd.ie)

Published in partnership with FCT NOVA with the support of E-MRS

npj nature partner journals

Although some progress has recently been made towards unravelling the physical origins of the LL-STEM-EEL features of $MoS_2$ as shown by Tizei, Lin[20, 21] significant gaps remain in our understanding of the observed signals and their origins. These purely experimental works focus on the detection and shifts of spectral features observed in EELS around 2 eV, which are presumed to be excitonic; however, no complementary theoretical work has been provided for the EELS of 2D $MoS_2$ to back up these assumptions. Here, we discuss the very first systematic study of all of the LL-STEM-EEL signals detected including their origins and local variations in mechanically exfoliated $MoS_2$, using a combination of state-of-the-art monochromated aberration-corrected dedicated STEM and ab-initio simulations. We observe large spatial variations of both exciton and plasmon signatures in the spectrum. In particular, we find that the two types of resonances behave qualitatively differently with respect to flake thickness and the position of the electron beam relative to the edge region. Based on ab-initio calculations we explain these effects in terms of the different dependence of the two types of excitations on the in-plane momentum transfer and the size of the dielectric constant. Also, we show that while the EELS and absorption spectra are generally different for bulk (three-dimensional (3D) materials), this is not the case in 2D materials for which the EELS and absorption spectra will be identical in the long-wavelength limit.

## RESULTS AND DISCUSSION

In Fig. 1a, b the integrated intensity maps of the LL-STEM-EEL signals from an edge region and a folded region of a typical $MoS_2$ flake are shown. Each pixel in these maps represents the intensity of an individual EEL spectrum integrated over a specific energy-loss window. It can be seen that the signal intensities vary largely between different regions of the samples. Some regions appear to 'light up' at certain energy-losses, representing an increase in intensity of these particular energy-loss peaks. Furthermore, the intensity maps generated from the energy-loss integration windows 1.8–2.2 and 3–3.5 eV in Fig. 1a (arrows) show an enhanced intensity on the edge region of the flake.

Low-loss EELS transitions are known to be delocalised, thus limiting the spatial resolution of plasmon mapping experiments to a range of a few nm, even though the electron probe used here is only ~ 1 Å in size.[22] There have however been recent reports of localised, sub-nm signal enhancement in the low-loss for specific structures such as the edge of a graphene sheet;[23] and $MoS_2$ nanostructures were recently predicted to support one-dimensional edge plasmons.[24] Our observations could thus be an indication of a distinct LL-STEM-EELS signature from the edge region. However, in the case of a true edge effect, no enhancement in intensity would be observed in the folded region for the same energy-loss integration windows (Fig. 1b, arrows). The observation of this increase in intensity of the signal at the fold suggests that the interpretation is less straightforward and that this effect may result from or be affected by different factors such as the geometry of the set-up. This highlights the need for a careful comparison of experiment and simulations in more controlled geometries, using a flake with stepped edges to assess the effects of edges, steps as well as the beam geometry onto the LL-STEM-EEL signals.

Local variations in experimental EEL spectra
The atomic-resolution high-angle annular-dark-field (HAADF) STEM image shown in Fig. 2a represents a typical edge region of a mechanically exfoliated $MoS_2$ flake as studied. The HAADF STEM image in Fig. 2b shows the region (boxed area) where the EELS analysis described in the following was recorded. The LL-STEM-EEL spectra were summed horizontally across the areas shown in Fig. 2b starting from the region over the vacuum (black, corresponding to spectrum 1 in Fig. 2d) to the innermost part of the $MoS_2$ flake (grey, spectrum 12). The intensity map in Fig. 2c shows the EEL signal integrated over a 19.7–20.7 eV energy window: using this energy-loss range, known to correspond to the bulk plasmon peak of $MoS_2$, the signal increases with each terrace and can, therefore, be used to illustrate the variations in thickness of the flake (black = vacuum, from dark blue to white = increasing thickness of the material). This area of the flake was determined to be 5 layers thick at the first terrace (T1), and 8 layers at the second terrace (T2), 10 layers for the third terrace (T3) and >10 layers in the thickest region (T4) (details of the thickness measurements can be found in the Supplementary Materials section S1). When comparing the LL-STEM-EELS of the edge, terraces, steps and centre of the flake regions, aside from the obvious thickness increase, the most prominent local variations observed were several features of interest at ~ 2 eV (labelled i), at ~ 3 eV (labelled ii and iii), at ~ 8.3 eV (labelled iv) and at ~ 14 eV (labelled v) as shown in Fig. 2d, which will be examined in greater detail in what follows.

Identifying and mapping excitons
Closer inspection of the spectral feature at ~ 2 eV (Fig. 2d, peak i) reveals that it consists of two sub-peaks centred at 1.88 and 2.08 eV, respectively. We postulate that these peaks correspond to known A and B excitonic transitions, that give rise to absorption peaks at 1.85 and 1.98 eV (ref. [17]), previously observed using optical methods[15, 24, 25] as well as LL-STEM-EELS.[20] The A and B excitons are formed by single-particle transitions from the spin-orbit split valence band at the K-point of the Brillouin zone (BZ).[26] The intensity and exact energy of these transitions are known to depend on temperature: our experiments were carried out at room temperature, and while earlier reports suggest that a lower observation temperature may be more favourable to detect these features in LL-STEM-EELS,[21] the very clear signal observed here and the split nature of the peak, strongly support their assignment as A and B excitons. Further discussion on the effects of temperature can be found in the Supplementary Materials section S2. Neither peak exhibited a detectable energy shift when comparing the signal with increasing flake thickness, which is in agreement with previous optical measurements where only small dependence of layer thickness on the peak position was found[15] (more discussion in the Supplementary Materials section S2 and refs [27, 28]) By contrast, the intensity of the signal was found to depend sensitively on the number of layers: energy-loss maps for integration windows at 1.8–2 and 2–2.2 eV (Fig. 3b—the spatial dependence of the bulk plasmon is also displayed for completeness) both reveal an enhanced signal on the thinnest region of the flake (T1). The intensity is reduced at T2 and even more at T3 and T4, which are the thicker regions of the flake. This suggests that the nature of this spectral feature is quite different from the bulk plasmon peak that grows in intensity towards the thicker regions of the flake. Also in the vacuum region the resonances show different behaviour (energy-losses can occur even when the beam does not traverse the sample directly: this corresponds to the so-called 'aloof' mode, which in a dielectric formulation can be thought of as arising from the interaction between the electric field generated by the beam electrons and the sample potential). For the bulk plasmon peak, the signal was found to decay to background level at position close to the edge of the flake (position 100 nm, Fig. 3b). In contrast, the signal for both excitonic peaks decayed to zero only at ~ 50 nm away from the edge (position 150 nm). We shall return to this observation later.

In order to support the interpretation of the different features in the loss-spectra, ab-initio calculations were performed for mono and few-layer $MoS_2$. The spatially resolved EEL spectra of a 2D material can be expressed in terms of the density-response-

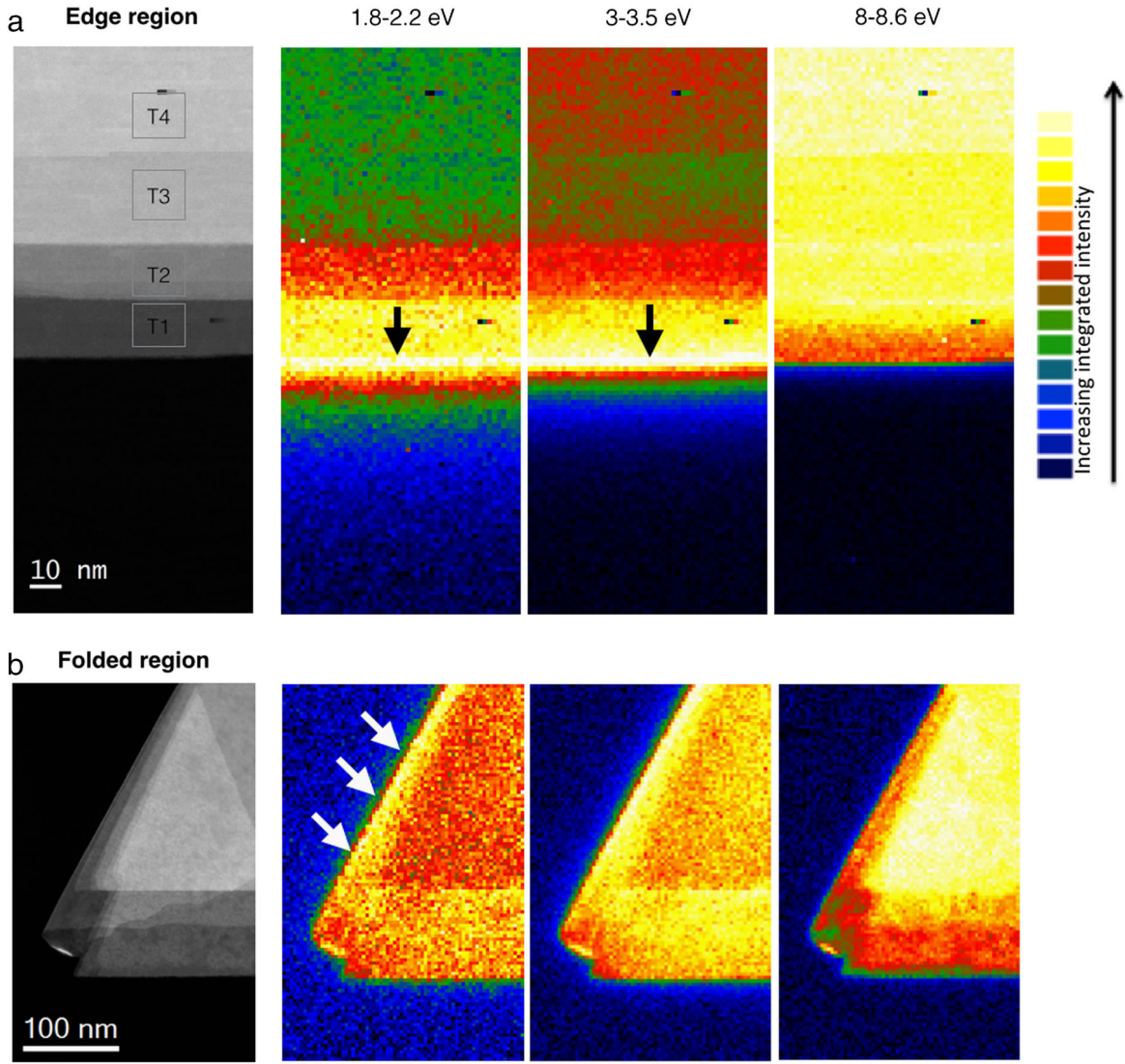

**Fig. 1** Experimental study of intensity variations in LL EELS maps acquired of flake edges, steps and folded regions of a MoS$_2$ flake. **a** HAADF STEM image and maps show the integrated intensity for energy-loss (EL) intensity integration windows 1.8–2.2, 3–3.5 and 8–8.6 eV of an edge region as well as **b** a folded region for comparison. The intensity maps in **a** show an enhanced intensity in the edge region of the flake (*arrows*) for EL 1.8–2.2 and 3–3.5 eV, suggesting the presence of an edge effect. However, the signal is also enhanced in the region of the folded area in **b** (*arrows*) for the same EL ranges. This shows that the enhanced signal at the edge is most likely an effect of the beam geometry and not an indication of a distinct LL-STEM-EELS signature from the edge region, a so-called 'edge effect'. For EL range 1.8–2.2 and 3–3.5 eV it was found that the signal intensities were the strongest in the thinnest region, T1, for the edge region as well as in the thinner regions of the folded flake. When comparing the signal from the regions T2 and T3, it was found that the signal slightly decreased with increasing thickness. In contrast, the signal in EL range 3–3.5 eV maps for both, edge and folded regions appear to increase in intensity with increasing thickness. The variations observed in the EELS maps for edge and folded regions highlight the need for an in-depth study of the LL-STEM-EELS as well as the effect of beam geometry on the signal

function, $\chi$ and the external potential of the electron beam, $V_{ext}$:

$$\text{EELS}(\omega) \propto -\text{Im}\sum_{q_\parallel=0}^{q_\parallel^{max}} V^*_{ext}(q_\parallel,\omega)\chi(q_\parallel,\omega)V_{ext}(q_\parallel,\omega), \quad (1)$$

(see methods and SI5 for more details). The local spectra generally involve a range of in-plane momentum-transfers. Due to the high velocity of the beam electrons, however, only negligible out-of-plane momentum transfers are involved and are set to zero in Eq. (1) for simplicity. The density-response-function and loss-spectra were calculated using two different approximations: one solving the Bethe-Salpeter equation (BSE), which includes electron–hole interactions but is computationally costly, and the other, based on the simpler time-dependent density-functional-theory (TDDFT) within the random phase approximation (RPA) without electron-hole coupling. In order to capture excitonic effects at low energies, the BSE must be applied, while plasmons are generally well described within the RPA.

Fig. 3c compares the BSE loss spectrum (blue) of monolayer MoS$_2$ to the RPA spectrum (green dashed). Both spectra are calculated on top of single-particle energies obtained from the $G_0W_0$ self-energy method.[29] The BSE spectrum clearly reveals the A and B excitonic peaks at ~2.1 eV, which exhibit a spin-orbit splitting of 0.15 eV, in good agreement with the experiment. In contrast, the RPA spectrum shows no features below the $G_0W_0$ band gap of 2.75 eV, confirming that the peaks observed experimentally indeed have excitonic character. Moreover, we found that the loss-spectra below 5 eV were almost completely

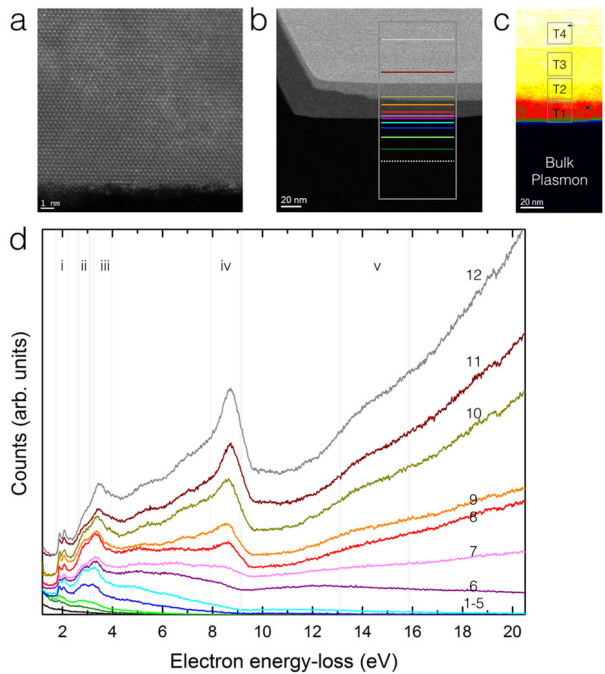

**Fig. 2** Scanning transmission electron microscopy (STEM) imaging and low-loss (LL) STEM electron-energy-loss (EEL) spectroscopy analysis of $MoS_2$ from the edge to the inside of flake of increasing thickness. **a** High-resolution high-angle annular dark field (HAADF) STEM image showing the edge region of a mechanically exfoliated $MoS_2$ flake. **b** HAADF STEM image showing the region of interest (ROI) (*boxed area*) of which monochromated LL-STEM-EELS were acquired. The edge of the flake as well as several steps were present in the ROI. **c** Integrated intensity map of the bulk plasmon peak (19.7–20.7 eV integration window) of the ROI showing the thickness variation (*black* = vacuum, from *dark blue* to *white* = increasing thickness of the material). The thinnest terrace is annotated T1, the second terrace annotated T2 and the thickest region is annotated T3; they were found to be 5, 8 and 10 layers thick, respectively. **d** Monochromated LL-STEM-EELS from the different areas as shown in **b**. The spectra shown were obtained from vacuum (black spectrum 1), aloof mode (spectra 2–4), the edge region (spectra 5–7), the first terrace T1 (spectrum 8), the step region to the second terrace T2 (spectrum 9), terrace T2 (spectrum 10), terrace 3 (spectrum 11) and terrace T4 (spectrum 12)

dominated by the small $q$ contributions and thus only the $q = 0$ spectra are shown in Fig. 3c.

In addition to the A and B excitons, the BSE spectra reveal a strong, split-nature, resonance at ~3 eV with a strong feature at ~2.8 eV and a tail extending to 3.5 eV, which is clearly downshifted in energy compared to the corresponding strong RPA@$G_0W_0$ peak at 3.5–4.8 eV. This resonance is also known as the C-exciton that was reported recently by Mertens, Shi,[30] and originates from transitions at the Γ-point of the BZ. The experimental spectra do indeed contain a split-nature resonance at around ~3 eV, labelled peaks ii and iii in Fig. 4a. The monochromated LL-STEM-EELS allowed for resolving two distinct peaks: a shoulder, centred at ~2.85 eV, as well as a strongly pronounced peak whose maximum shifts from 3.2 to 3.5 eV with increasing thickness (Fig. 4b). As was the case for the excitonic peak(s) at ~2eV, this loss feature also showed enhanced intensity at the thinnest part of the flake (the first terrace, labelled T1), with a noticeable enhancement at the edge of the flake. This is clearly illustrated in the map generated by integrating the intensity over a 3–3.5 eV window, but also when mapping the intensity of the tracked maximum of peaks ii and iii individually (Fig. 4c). As for peak i, peaks ii and iii show significant intensity even beyond the edge of the flake (Fig. 4c, dashed red line). Thus, the intensity of the ~3 eV peaks behaves in a way analogous to that of the A and B exciton. Together with the presence of the split ~3 eV peak in the theoretical BSE spectrum, this suggests that peaks ii and iii are of an excitonic nature and we propose that they correspond to the C-exciton of $MoS_2$. We find that the splitting of this peak increases with number of layers.

Identifying and mapping plasmons

We now turn to the discussion of the spectral features at energy above 8 eV. Firstly, the experimental spectra reveal an increase in intensity with increasing number of layers for the higher energy-loss peaks: for the well-documented bulk plasmon peak as well as for peak iv, as shown in the maps of integration window 19.7–20.7 eV (Fig. 2c) and 8–8.6 eV (Fig. 1a), respectively. This behaviour is also observed in the individual experimental EEL spectra (Fig. 2d) for the bulk plasmon peak and peak iv.

In order to inform the interpretation of the higher energy-loss peaks, theoretical EEL spectra were calculated using the RPA approach. The lower computational cost of RPA allows us to perform full ab-initio calculations for 1–3 layers of $MoS_2$, while a recently developed multi-scale approach, the quantum electro-static heterostructure (QEH) model[31] was used to obtain the spectra for up to 10 layers. The calculated RPA loss-spectra for 1–3 layers of freestanding $MoS_2$ are shown in Fig. 5a. The QEH result (dashed lines) is found to be in good agreement with the full ab-initio approach (solid lines). The spectra are resolved with respect to in-plane momentum transfer $q_{||}$ (perpendicular to the incident beam direction), which have the largest contribution to the experimentally recorded spectra. Here, we identify electron-loss peaks at ~3, ~8.3, and ~14 eV, where the ~3 eV peak is seen to have a large weight only for $q_{||} = 0$ (as previously noted), while the ~8.3, and ~14 eV peaks appear only for $q_{||} > 0$.

The effect of number of layers/thickness on spectra

The large difference between the $q_{||} = 0$ result and the spectra for finite $q_{||}$, can be understood by considering the Dyson equation for the density-response-function,

$$\chi(q_{||}, \omega) = \chi^{irr}(q_{||}, \omega) + \chi^{irr}(q_{||}, \omega)V(q_{||})\chi(q_{||}, \omega), \quad (2)$$

where $\chi^{irr}(q_{||}, \omega)$ is the irreducible density-response-function. Within RPA $\chi^{irr} = \chi^0$, i.e., the single-particle response. For a semiconductor such as $MoS_2$, $\chi^{irr} \propto q_{||}^2$ for small $q$. Since the Fourier transformed Coulomb potential, $V(q)$, takes the form $1/q$ in 2D, this implies that the long-ranged coupling governed by the Coulomb interaction (second term on the right-hand-side of Eq. (2)) is effectively turned off as $q \to 0$. Consequently, plasmons do not form in this limit and the EEL spectra for $q = 0$ can simply be thought of as the material's absorption function. This is qualitatively different from the behaviour of 3D materials, where the EELS and absorption spectra are generally different. Therefore, in the RPA where the ~3 eV peak originates from single-particle transitions, the simulated intensity is simply found to scale with the number of layers $N = (1,2,3)$. This is in contrast to the experiments that find a reduced intensity of the lower energy-loss peaks on the thicker parts of the sample. We ascribe this to the weakening of the electron-hole interaction and thus the excitonic effects due to the enhanced screening in thicker layers. This effect is not captured by the RPA calculations. However, using BSE calculations we observed a decrease in intensity as well a broadening of the excitonic features at ~2 and ~3 eV with increasing N (Supplementary Materials section S3).

Contrary to the ~3 eV peak, the calculated peaks at ~8.3 and ~14 eV become more prominent for non-zero $q_{||}$. Furthermore, the peaks shift to higher energy losses with increasing $N$ and increasing $q_{||}$. This is a clear signature of plasmonic character, since (1) the electrostatic coupling between plasmons in different

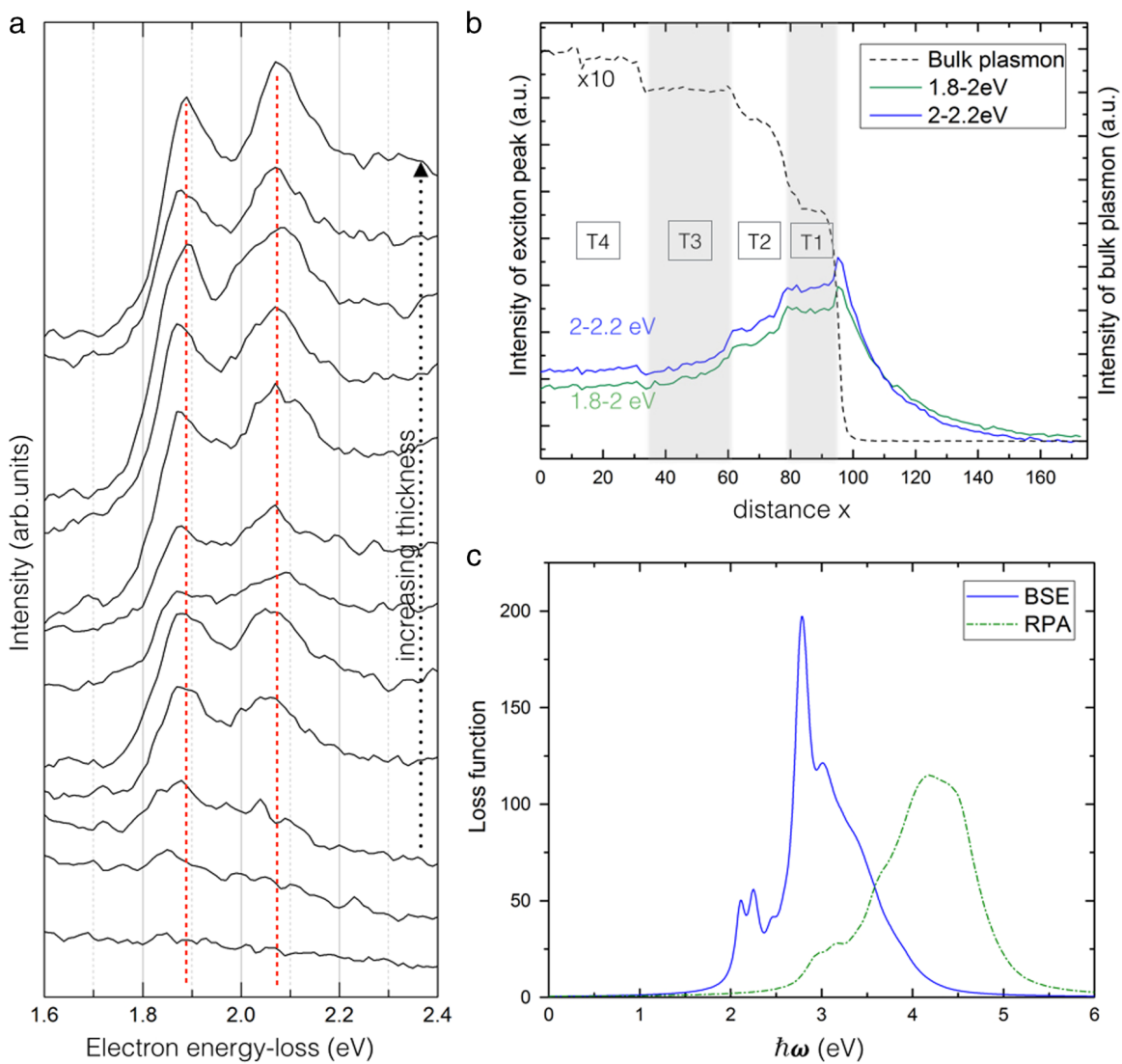

**Fig. 3** Excitonic origin of ~2 eV peaks shown using Bethe-Salpeter equation (BSE) and monochromated LL-STEM-EELS. **a** Details from the monochromated LL-STEM-EELS (as found in Fig. 2d, peak i), showing the splitting of the peak—one peak is centred at ~1.88 eV and one at ~2.08 eV. No peak shift was observed of either peak when comparing edge regions with the inside of the flakes. **b** Line profiles obtained showing the intensity variation from the inside of the flake to the vacuum of the LL-STEM-EELS for energy-loss integration windows 1.8–2 and 2–2.2 eV as well as 19.7–20.7 eV (bulk plasmon peak region) from the region as shown in Fig. 2b. For the bulk plasmon peak, the signal decays to 0 at position x = 100 nm, close to the edge of the flake. In contrast, the signals for 1.8–2 and 2–2.2 eV decay to 0 at x~ 150 nm only. **c** Results from the BSE (*blue, solid line*) and RPA calculations (*green dashed line*) for a monolayer $MoS_2$. RPA represents a quasi-particle spectrum without the inclusion of electron-hole coupling. This gives excitonic peaks at ~2.1 eV—exhibiting A-B splitting of ~0.15 eV. This is in close agreement with the experiments where excitonic peaks were found at ~1.88 and ~2.08 eV. In addition, there is a resonance at ~3 eV. Overall, there is a large difference between the BSE and the RPA spectra, hence the excitonic effects which are at the origin of this difference are large. Only vanishingly small effects were found when including higher q contributions in the spectra, q = 0 dominates the spectra for excitons

layers (also known as plasmon hybridisation) will lead to a blueshift of plasmon energy, and (2) plasmons generally show a strong (positive) energy dispersion with q. Also, a blueshift of plasmon resonances with increasing N were previously observed in several studies of graphene.[32–34] We propose that the TDDFT calculated peaks at ~8.3 eV (Fig. 5a, b; peak iv), and at ~14 eV (Fig. 5a, b; peak v) correspond to the peaks found experimentally at ~8.3 eV (Fig. 2d, peak iv) and at ~14 eV (Fig. 2d, peak v), respectively. Thus, the calculations strongly supports that the peaks at ~8.3 and ~14 eV have plasmonic origin.

According to the above theoretical analysis, plasmons should dominate the loss spectrum for larger $q_\parallel$ while excitons should dominate for small $q_\parallel$. This is supported by the experimental observation that the intensity of the high energy features (peaks iv and v) vanishes quickly when the electron beam passes the edge region into the vacuum while the signal from the low energy excitonic features (peaks i, ii and iii) decay slowly (see maps in Fig. 1a). Indeed, as the beam is moved further away from the sample, only the smaller $q_\parallel$-components of the 1/r potential from the electron beam will reach the sample thus effectively probing excitations with lower and lower in-plane momentum.

The strong q-dependence of the plasmonic features of the spectra means that we must take an average over amplitudes of in-plane momentum transfers (see Eq. 1) in order to produce local EEL spectra that are directly comparable to experiments. The loss-spectra for monolayer $MoS_2$, summed over $q_\parallel$ contributions is shown in Fig. 5b where it is compared to the $q_\parallel = 0$ spectrum. There is a clear difference between the $q_\parallel = 0$ and $q_\parallel$-summed

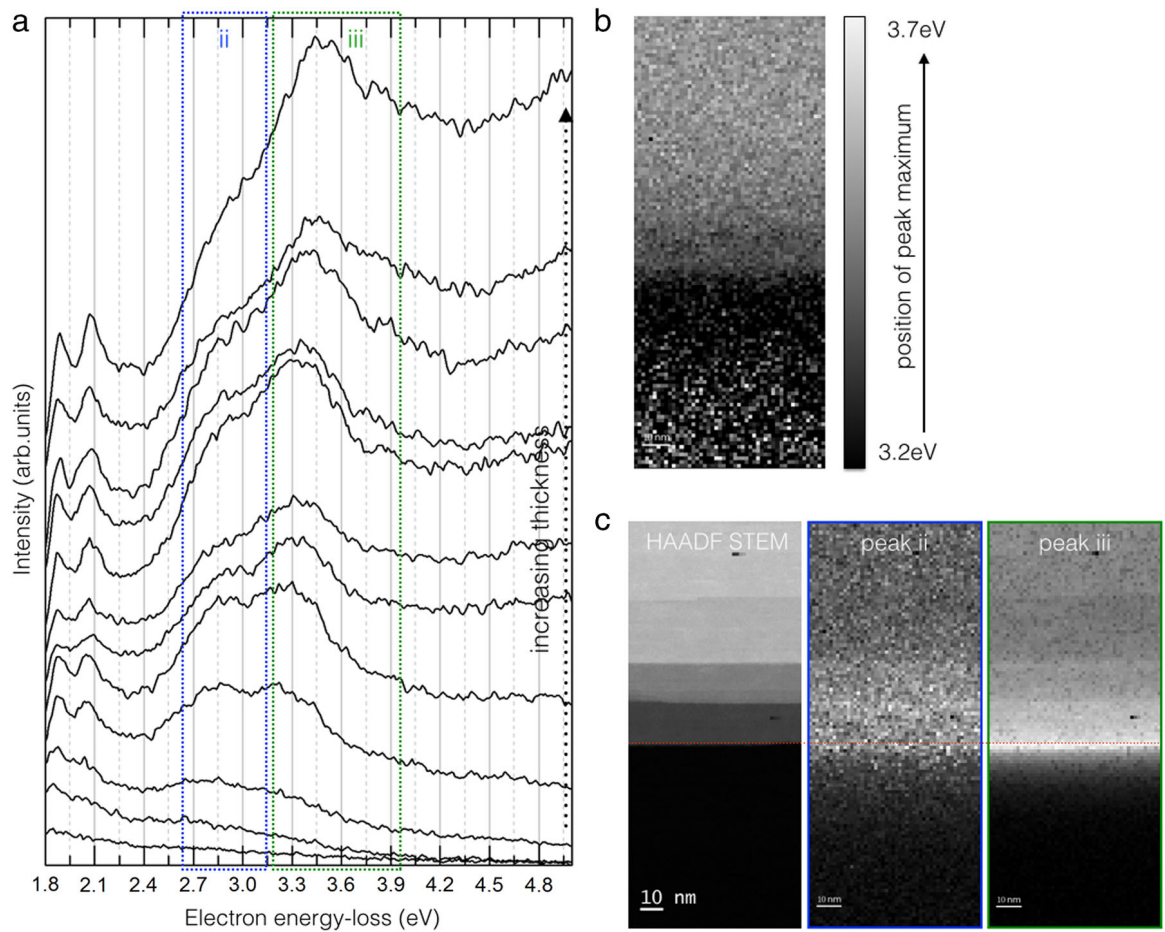

**Fig. 4** Shifts and intensity variations of peaks at ~ 3 eV. **a** Details from EEL spectra as shown in Fig. 2d, peaks ii and iii. When studying the monochromated LL-STEM-EELS at the ~ 3 eV energy-loss, it was found that we are observing two distinct peaks- one centred at around 2.85 eV and one peak with a maximum shifting from about ~ 3.2 to 3.5 eV when comparing the spectra from the edge (spectra 4 and 5) to the inside of the flakes (spectra 12, top spectrum here). **b** Map of the peak position of the higher energy-loss feature, as found by curve fitting to the peak maxima for all the peaks between 1.5 and 3.8 eV and tracking the position of the ~ 3 eV feature between 3–3.8 eV (scale: from *black* = 3.2 eV to *white* = 3.7 eV). **c** HAADF STEM image and map of the peak intensity of the 2.85 eV feature (peak ii) and the 3.2–3.8 eV feature (peak iii, tracked) as found by curve fitting to the peak maxima for all the peaks between 1.5 eV and 3.8 eV (intensity scale=arbitrary units). Both maps showed slightly enhanced intensity in the region of the first and the second terrace, the thinner region of the flake. In addition, the intensity of the 3.2–3.8 eV feature (peak iii, tracked) was found to be greatly enhanced at the edge of the flake as well as slightly enhanced at the step. Qualitatively, both peaks ii and iii follow a similar intensity distribution

spectra, where the ~ 8.3 and ~ 14 eV peaks become prominent in the latter. This provides a close correspondence between calculations and experiments (compared in Fig. 5b, c), in particular for the peak at ~ 8.3 eV (peak iv). In order to obtain a closer resemblance of the experimental set-up, the QEH approach was used to calculate $q$-summed spectra for up to 10 layers (Fig. 5d). The theoretical spectra clearly reproduce a plasmonic 8.3 eV peak, that is slightly blueshifted with increasing N, and a ~ 14 eV peak that strongly blueshifts towards a broad resonance from 14–28 eV for 10 layers.

The effect of beam geometry on spectra
Furthermore, the strong $q$-dependence of the EEL spectra implies that the geometry of the electron beam could significantly affect the outcome of the measurements, since this determines the range of momentum transfers that is probed. When comparing the interaction of the electron beam with the MoS$_2$ flake at its edge and within its inner regions of increasing thickness, there is a change in relative sample-beam geometry. This change results in a different angular range of $q$ being probed depending on the beam position leading to differences in the relative contribution of energy-losses to the spectrum with in- and out-of-plane momentum transfer. In turn, the TDDFT calculations for different values of $q_∥$ shown in Fig. 5 support this interpretation: as the geometry changes, the range and relative weight of momentum transfers over which the integration is carried out is affected, resulting in some of the observed variations in the spectra. In addition, we found that the weight of the $q = 0$ contribution increases with increasing accelerating voltage (Supplementary Materials section S4), since a different range of $q_∥$ is sampled as the electrons move at different speeds through the sample.

Earlier plasmonic studies of 2D materials[35, 36] suggested that the number of layers (and, therefore, the thickness) of the flakes could be directly 'fingerprinted' through LL-STEM-EEL spectrum. Our combination of experimental observations and careful ab-initio calculations, suggests that while this is true, the effect is due to beam/sample geometry effects (and, therefore, very dependent upon the exact optical parameters chosen for the experiments), rather than a true reflection of drastic changes in electronic structure and properties between 'real' 2D and thin bulk (3D).

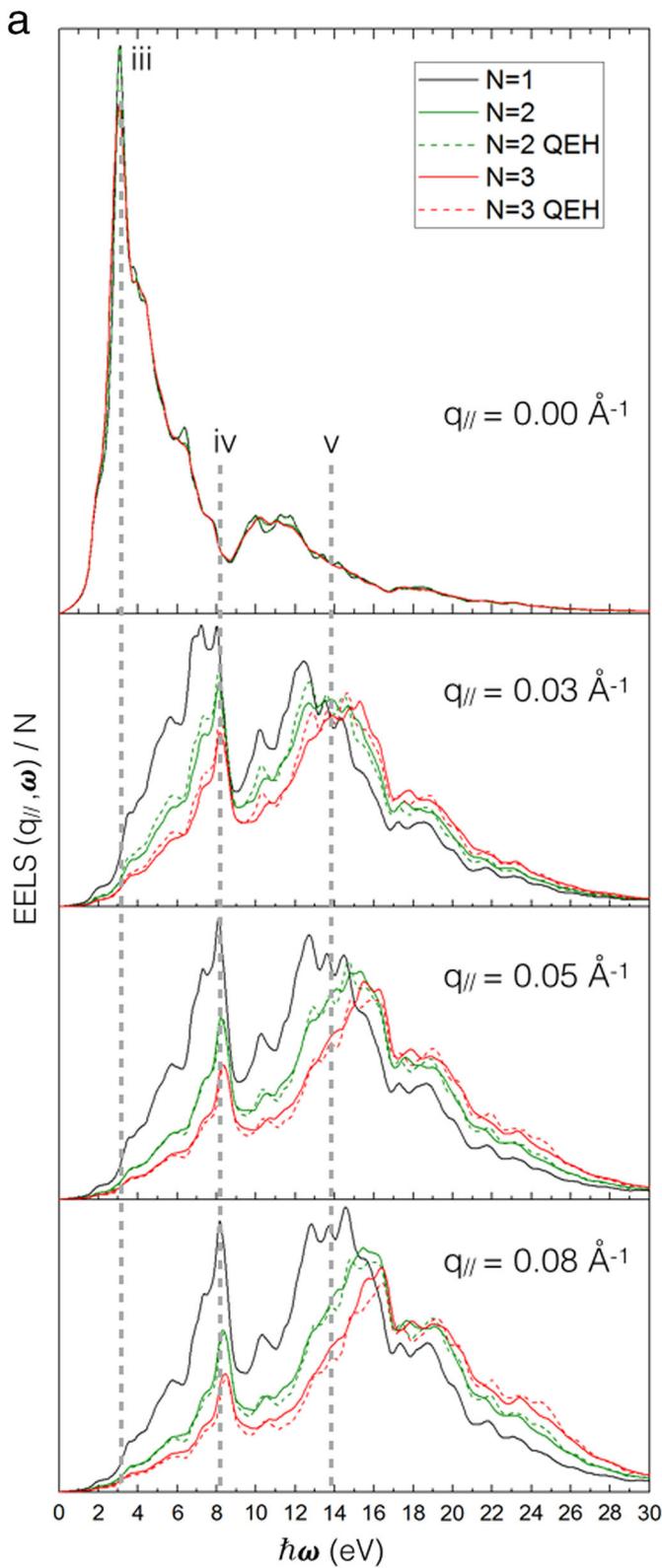
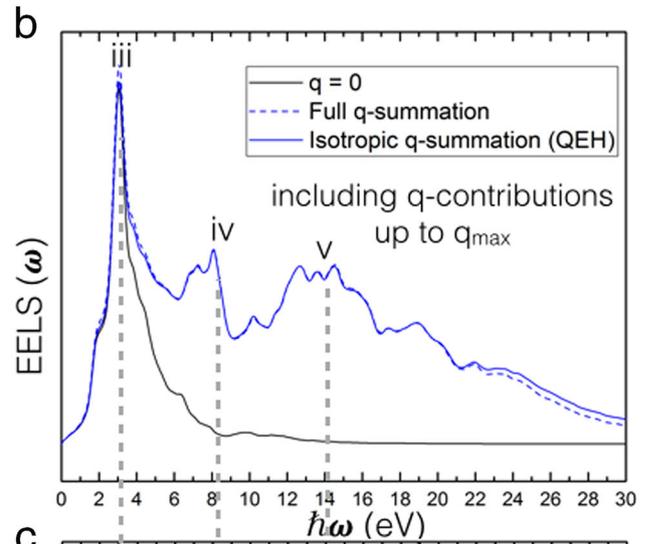
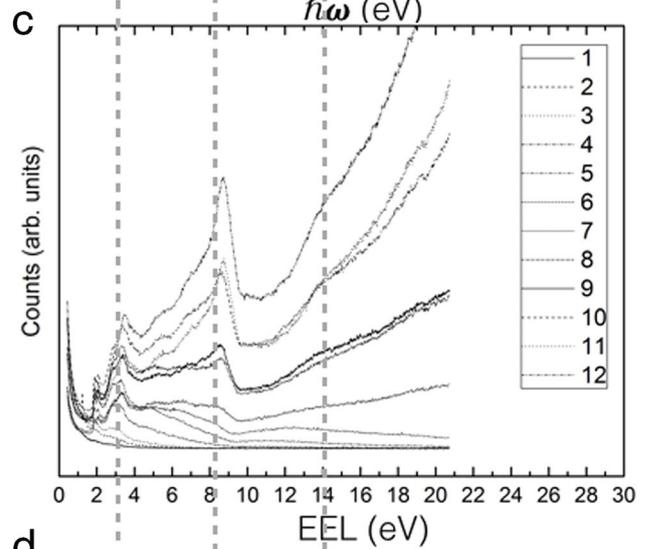
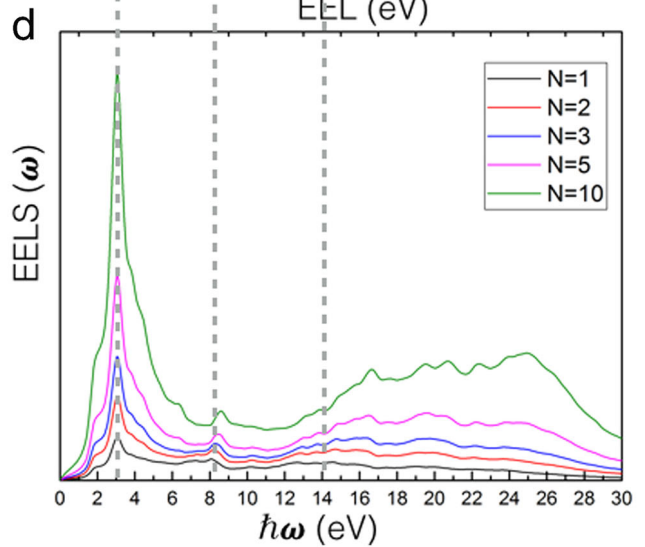

## CONCLUSION

In conclusion, using LL-STEM-EELS and many-body calculations we have presented a combined spatial and energy-resolved analysis of the dominating collective excitations in few-layer $MoS_2$. Whereas previous studies using LL-STEM-EELS were purely experimental,[20, 21] we use here a combination of experimental and several computational techniques which allow us to compare directly experimental and calculated spectra, and thus assign unambiguously each experimentally observed peak to a specific type of excitation. In particular, we show that it is possible to detect the presence of the C exciton in our data. Using BSE calculations we confirmed the excitonic nature of the peaks at ~ 2 eV, also known as the A and B excitons. Furthermore, we note that the application of BSE methods to LL-STEM-EELS calculations is very much in its infancy, and this is to our knowledge the first direct confirmation of the excitonic nature of the peaks observed in STEM-EELS through ab initio simulations.

Fig. 5 Effect of electron beam geometry and number of layers studied using time-dependent density-functional-theory (TDDFT) in the random phase approximation (RPA) and compared to experimental monochromated LL-STEM-EELS. **a** EEL spectra of 1–3 layers of freestanding $MoS_2$ found using RPA. The $q$-resolved RPA spectra shown here are calculated for the $q_\parallel$ in the Γ-M direction of the 2D Brillouin zone and normalised to the number of layers. RPA gives a good description of plasmons but not of excitons due to exclusion of electron–hole interactions, essential for excitons. For $q_\parallel = 0$ no difference was found with increasing $N$. Once the corrections are applied, the energy-loss is found to set in at ~ 3 eV (iii). For larger $q$, however, the spectra diverge for increasing $N$. Hence for non-zero $q$, peaks associated with plasmons come in at ~ 8.3 eV (iv) and ~ 14 eV (v). The peaks in the spectra associated with the plasmons also blueshift. Hence, the differences in the spectra with increasing number of layers as observed experimentally are due to beam geometry. The beam is probing a large range of $q$ in plane across the sample, however, as the structure becomes 3D (with increasing number of layers) the beam is also increasingly probing out-of-plane excitations and as a consequence, the shape of experimental spectra changes with increasing number of layers. **b** The spectra summed over $q$ contributions is shown here with $q_{max} = 2.0 A^{-1}$. When comparing the spectrum for $q = 0$ and the spectrum for summed $q$ up to $q_{max}$, there is a peak at ~ 8.3 eV (iv) as well as at ~ 14 eV (v) that become prominent in the latter spectra. When compared to experimentally acquired spectra as shown in **c**, the peaks at ~ 3 eV (peak iii), the peak at ~ 8.3 eV (peak v) as well as the peak at ~ 14 eV (peak v) show a nice correspondence between experiment and calculations. In addition, the ~ 3 eV peak (iii) is present in all of the experimental as well as calculated spectra. The peak at ~ 8.3 eV is present in our calculated spectra for finite $q$ as well as in some of the experimental spectra, where the peak appears and increases with increasing number of layers, supporting our earlier hypothesis about the beam probing an increasing contribution from $q$ out-of-plane with increasing thickness (increasing number of layers N). **d** The spectra were also calculated for increasing number of layers N (for N = 1,2,3,5). The contribution from $q = 0$ spectrum to the summed spectra grows proportional with N. The plasmon peaks were also found to blueshift with increasing N

In addition, we showed that the excitonic and plasmonic features in the loss spectrum behave qualitatively differently both with respect to the number of layers and the in-plane momentum transfer. Whereas previous studies on graphene[32, 35, 36] suggested that the peak positions as well as the presence/absence of the peaks in the plasmon region of the LL-STEM-EELS could be used to determine the exact number of layers, we showed in this study that the interpretation of the LL-STEM-EELS is not as straight forward. We rationalise through a comparison with theoretical calculations the dependence of the experimental spectra on beam geometry (through an integration over momentum space, to take into account the beam convergence), and on sample geometry (number of layers), demonstrating a good agreement between theory and experiment. This in turn suggests that while the plasmon region of the low loss spectra of 2D materials is sensitive to the number of layers, great care must be taken to use this shape and intensity dependence as a direct fingerprint of the number of layers, as complex simulations are necessary to fully account for all the effects that are at play. The results presented here open a new avenue for understanding the physics of plasmons and excitons in nanostructured materials.

## METHODS

### Materials
Bulk crystal $MoS_2$ was purchased from SPI Supplies and the mechanically cleaved samples of $MoS_2$ were produced following the procedure as described by Novoselov et al.[3] and then transferred onto SPI holey Carbon finder grids.

### Characterisation and equipment
STEM imaging and LL-STEM-EELS of the mechanically cleaved samples was mostly carried out on the monochromated Nion UltraSTEM 100MC 'HERMES' fitted with a C5 Nion QO corrector and equipped with a Gatan Enfinium ERS spectrometer (SuperSTEM, UK). The microscope was operated at an acceleration voltage of 60 kV to minimise knock-on radiation damage. EELS were acquired with a convergence semi-angle of 34 mrad, a collection semi-angle of approximately 44 mrad, a dispersion of 0.01 eV/channel and an energy resolution over vacuum of at least 0.04 eV as estimated from the zero-loss peak (ZLP) full-width at half-maximum. LL-STEM-EELS maps were acquired with 0.01 s acquisition time per spectra per pixel. After acquisition, dark reference correction was applied to the spectra and the spectra were aligned and calibrated using the ZLP. The spectra shown here are representative sets of a minimum of 10 LL-STEM-EELS maps that were acquired over similar regions. For Figs 2 and 3, the EEL spectra as shown were summed horizontally across the whole region shown in the HAADF STEM image in Fig. 2b, before line profiles (Fig. 3b) were obtained of the intensity starting inside the flake and extending into the vacuum region. Artefacts due to Cerenkov radiation and retardation effects were minimised here due to the use of a 60 kV beam and low sample thickness.[19] Hence any intensity below the band gap energy was assumed to arise from another source.

### Processing
Processing was kept to a minimum due to the delicate nature of the LL-STEM-EELS analysis. No spectral de-convolution was necessary since the overlap of the signal with the ZLP was not significant. The experimental spectra (when displayed as spectra, not as maps) as shown in Figs 2–5 were normalised to have the same total counts and spread out in the y direction for clarity of display.

A fitting procedure using hyperspy[37] to fit all the peaks between 1.5 and 4 eV was applied to track the position of the maxima as well as the intensity of the peaks as shown in Fig. 4.

### Computational work
Two computational approaches were combined here to study the plasmonic as well as the excitonic signatures of $MoS_2$.

The plasmonic properties were studied with linear response TDDFT,[38] where the electronic density-response-function was calculated on top of the Kohn-Sham states obtained from a DFT calculation, as implemented in the electronic structure code GPAW.[39] The RPA was used to calculate the interacting density-response-function of the material:

$$\chi(\mathbf{r},\mathbf{r}',\omega) = \chi^0(\mathbf{r},\mathbf{r}',\omega) + \iint d\mathbf{r}_1 d\mathbf{r}_2 \chi^0(\mathbf{r},\mathbf{r}_1,\omega) \frac{1}{|\mathbf{r}_1 - \mathbf{r}_2'|} \chi(\mathbf{r}_2,\mathbf{r}',\omega)$$

where only the Coulomb potential, $V_C = \frac{1}{|\mathbf{r}_1 - \mathbf{r}_2'|}$, is included in the Kernel.

In practice, the response functions are represented in reciprocal space, and are obtained as a function of the in-plane momentum transfer, $q_\parallel$, of the 2D layers. The response function is related to the $q$-resolved energy-loss-spectra, by:

$$EELS(q_\parallel,\omega) = \frac{4\pi}{q_\parallel^2} \text{Im}\chi(q_\parallel,\omega)$$

Full ab-initio calculations have been performed for 1, 2 and 3 layers of $MoS_2$, while a recently developed multi-scale approach, the QEH model,[31] has been used to calculate the response for up to 10 atomic layers. Within this framework, spectra summed over $q$-contributions are also calculated in order to obtain better comparison with experiments.

TDDFT within the RPA generally gives a good description of plasmons, but does not include the electron–hole interaction that is essential for accounting for excitonic effects. Instead, we employ the Bethe Salpeter Equation (BSE). Since this approach is computationally demanding, it is only applied for calculating the spectrum for a few eV above the bandgap of monolayer $MoS_2$. The BSE spectra were calculated on top of $G_0W_0$ quasiparticle energies,[29] where the spin-orbit coupling is included non-self-consistently. More calculational details are provided in the supplementary information (S5).

## ACKNOWLEDGEMENTS
We wish to acknowledge SuperSTEM, the U.K. National Facility for Aberration-Corrected STEM, supported by the Engineering and Physical Sciences Research Council (EPSRC). V.N. and H.C.N. wish to acknowledge the following funding support: SFI AMBER, SFI PIYRA, ERC StG 2DNanoCaps, ERC PoC 2DUSD, ERC PoC 2DInk, FP7 MC ITN MoWSeS, Horizon2020 NMP Co-Pilot. V.N. and H.C.N. also acknowledge CRANN's Advanced Microscopy laboratory in Trinity College Dublin for the experimental help and the provision of their facilities. The Center for Nanostructured Graphene (CNG) is sponsored by the Danish National Research Foundation, Project No. DNRF58.

## AUTHOR CONTRIBUTIONS
H.C.N., Q.M.R. and V.N. conceived the project. H.C.N., F.S.H. and Q.M.R. performed electron microscopy characterisation and analysis. K.T.W. and K.S.T. performed and analysed computational simulations. H.C.N. and L.H. carried out preliminary electron microscopy experiments. H.C.N., C.B. and J.N.C. were involved in preliminary experiments. H.C.N., K.T.W., K.S.T. and Q.M.R. co-wrote the paper with all authors contributing to the discussion and preparation of the manuscript.

## COMPETING INTERESTS
The authors declare that they have no competing interests.

Supplementary Information accompanies the paper on the *npj 2D Materials and Applications* website (doi:10.1038/s41699-017-0003-9).